\documentclass[aps,prl,twocolumn,citeautoscript,showkeys]{revtex4-1} 
\usepackage{graphicx}  % needed for figures
\usepackage{dcolumn}   % needed for some tables
\usepackage{bm}        % for math
\usepackage{amssymb}   % for math
\usepackage{epsfig}
\usepackage{epstopdf}  
\usepackage{subfigure}
\usepackage{tensor}
\usepackage[utf8]{inputenc}
\usepackage{mdframed}
\usepackage{amsmath}
\usepackage{protosem}
\usepackage{wasysym}
\usepackage[colorlinks=true]{hyperref}  
\hypersetup{
    bookmarks=true,         % show bookmarks bar?
    unicode=false,          % non-Latin characters 
    pdftoolbar=true,        % show Acrobat
    pdfmenubar=true,        % show Acrobat 
    pdffitwindow=false,     % window fit to page when opened
    pdfstartview={FitH},    % fits the width of the page to the window
    pdftitle={My title},    % title
    pdfauthor={Author},     % author
    pdfsubject={Subject},   % subject of the document
    pdfcreator={Creator},   % creator of the document
    pdfproducer={Producer}, % producer of the document
    pdfkeywords={keyword1} {key2} {key3}, % list of keywords
    pdfnewwindow=true,      % links in new window
    colorlinks=true,       % false: boxed links; true: colored links
    linkcolor=magenta, %red,          % color of internal links (change box color with linkbordercolor)
    citecolor=blue,        % color of links to bibliography
    filecolor=magenta,      % color of file links
    urlcolor=cyan           % color of external links
} 

\newcommand{\bea}{\begin{eqnarray}}
\newcommand{\eea}{\end{eqnarray}}
\newcommand{\ba}{\begin{eqnarray}}
\newcommand{\ea}{\end{eqnarray}}

\newcommand{\beq}{\begin{equation}}
\newcommand{\eeq}{\end{equation}}
\newcommand{\beqa}{\begin{eqnarray}}
\newcommand{\eeqa}{\end{eqnarray}}
\newcommand{\beqar}{\begin{eqnarray*}}
\newcommand{\eeqar}{\end{eqnarray*}}

\newcommand{\ssc}{\scriptscriptstyle}
\newcommand{\eg}{{\it e.g.,}\ }
\newcommand{\ie}{{\it i.e.,}\ }

\newcommand{\ctt}{C_{\ssc T}}

\newcommand{\req}[1]{(\ref{#1})} %{Eq.\thinspace(\ref{#1})}  

\begin{document}

\title{Einsteinian cubic gravity} 
\author{Pablo Bueno$^{\text{\lightning}}$ and Pablo A. Cano$^{\textproto{\Ahe}}$}
\affiliation{\vspace{0.1cm}$^{\text{\lightning}}$Instituut voor Theoretische Fysica, KU Leuven,
Celestijnenlaan 200D, B-3001 Leuven, Belgium\\
$^{ \textproto{\Ahe}}$Instituto de F\'isica Te\'orica UAM/CSIC,
C/ Nicol\'as Cabrera, 13-15, C.U. Cantoblanco, 28049 Madrid, Spain\vspace{0.1cm}}
\date{\today}

\begin{abstract}
We drastically simplify the problem of linearizing a general higher-order theory of gravity. We reduce it to the evaluation of its Lagrangian on a particular Riemann tensor depending on two parameters, and the computation of two derivatives with respect to one of those parameters. We use our method to construct a $D$-dimensional cubic theory of gravity which satisfies the following properties: 1) it shares the spectrum of Einstein gravity, \emph{i.e.}, it only propagates a transverse and massless graviton on a maximally symmetric background; 2) the relative coefficients of the different curvature invariants involved are the same in all dimensions; 3) it is neither trivial nor topological in four dimensions. Up to cubic order in curvature, the only previously known theories satisfying the first two requirements are the Lovelock ones: Einstein gravity, Gauss-Bonnet and cubic-Lovelock. Of course, the last two theories fail to satisfy requirement 3 as they are, respectively, topological and trivial in four dimensions. We show that, up to cubic order, there exists only one additional theory satisfying requirements 1 and 2. Interestingly, this theory is, along with Einstein gravity, the only theory which also satisfies 3.

\end{abstract}

\maketitle
%%%%%%%%%%%%%%%%%%%%%%%%%%%%%%%%%%%%%%%%%%%%%%%%%%%%%%%%%%%%%%%%%%%%%%
%%%%%%%%%%%%%%%%%%%%%%%%%%%%%%%%%%%%%%%%%%%%%%%%%%%%%%%%%%%%%%%%%%%%%%
%%%%%%%%%%%%%%%%%%%%%%%%%%%%%%%%%%%%%%%%%%%%%%%%%%%%%%%%%%%%%%%%%%%%%%
%%%%%%%%%%%%%%%%%%%%%%%%%%%%%%%%%%%%%%%%%%%%%%%%%%%%%%%%%%%%%%%%%%%%%%
Higher-order gravities play a prominent role in different areas of high-energy physics. In cosmology, they have been countlessly considered in the search for a coherent picture of the history of the universe which can account for the observational evidence currently associated to early-time inflation, late-time acceleration or dark matter --- see \eg \cite{Sotiriou,Nojiri:2006ri,Nojiri:2010wj,Clifton:2011jh}. In holography \cite{Maldacena,Witten,Gubser}, they are often used to study different aspects of strongly coupled conformal field theories (CFTs) and, in some occasions, they have been crucial in unveiling certain universal properties of general CFTs --- see \eg \cite{Brigante:2007nu,Myers:2010tj,Myers:2010xs,Bueno1,Bueno2,Mezei:2014zla}. In fact, holography itself has motivated the construction of new higher-derivative theories like quasi-topological (QT) gravity \cite{Quasi,Quasi2,Myers:2010jv}. 

More broadly, higher-order corrections to the Einstein-Hilbert term should be produced in the gravitational effective action by the corresponding underlying ultraviolet-complete theory --- presumably String Theory \cite{Gross:1986mw,Green:2003an,Frolov:2001xr}. A more practical approach in this direction consists in considering certain classes of higher-order gravities as quantum gravity toy models \cite{Stelle:1976gc,Stelle:1977ry}. Popular examples of this are
topologically massive gravity \cite{Deser:1981wh} and new massive gravity \cite{Bergshoeff:2009hq} in three dimensions, and critical gravity in four \cite{Lu}. 

A particularly relevant aspect of a given higher-order gravity is its linearized spectrum, \ie the set of physical degrees of freedom propagated by metric perturbations on the vacuum. For example, in the context of holography, the linearized equations of a given higher-order gravity provide useful information about the corresponding holographic CFT stress-tensors, since these are dual to the metric perturbation --- see \eg \cite{Liu:1998bu,Myers:2010tj,Buchel:2009sk,Bueno2}.

%Generically, higher-order gravities incorporate, in addition to the usual massless graviton, a ghost-like massive graviton as well as a massive scalar field. As we will see, in some theories some of these modes are absent. %For example, f$($Lovelock$)$ is free of massive gravitons \cite{Love}, while conformal gravity does not incorporate the scalar mode \cite{conformal,Lu}.
As we will see, there are certain higher-order gravities which are equivalent to Einstein gravity at the linearized level in the vacuum, \ie the only physical mode propagated by the metric perturbation is a transverse and massless graviton. Some known examples include QT gravity \cite{Quasi}, and certain f$($Lovelock$)$ theories \cite{Love,Tekin3}. However, most of these theories have the inconvenience that the couplings of the different curvature invariants depend on the spacetime dimension $D$. Hence, they are actually different theories in different dimensions. The only known theories with dimension-independent couplings which share spectrum with Einstein gravity are Lovelock theories \cite{Lovelock1,Lovelock2,Padmanabhan:2013xyr}. 

We will show that, up to cubic order in curvature, there is one only additional theory which satisfies this criterium. Furthermore, as opposed to the quadratic and cubic Lovelock ones, this theory is non-trivial in four dimensions.

In the construction of this \emph{Einsteinian cubic gravity}, we make use of a new method for linearizing general higher-order gravities which we also present here. We will argue that this method is much faster than all the previous procedures available in the literature.

% \comment{ Some theories lack some of them: $f(R)$, $f(Love)$, quasi-topological, conformal gravity, critical gravity, etc. Focus on theories which have the same spectrum as Einstein. Some cases known, like Lovelock or quasi-topological. If you require them to be non-trivial in four dimensions, less options. Still some examples from f(Love). Only known theories which share spectrum with Einstein and have the same couplings in all dimensions are Lovelock ones. We show that up to cubic order, these are indeed all, except for a new one which, in addition, is non-trivial in four. \cite{Lu,Love,Quasi2,Quasi}}

\textbf{Linearization procedure:}
Let us consider a general $D$-dimensional theory of gravity involving arbitrary contractions of the Riemann tensor and the metric, \ie
\begin{equation}\label{Smass}
S=\int d^Dx\sqrt{-g}\, \mathcal{L}(R_{\mu\nu\rho\sigma},g_{\alpha\beta}) \, .
\end{equation}
The field equations of this theory can be written as
\begin{equation}
\mathcal{E}_{\mu\nu}=P_{\mu\sigma\rho\lambda}R_{\nu}\,^{\sigma \rho\lambda}-\frac{1}{2}g_{\mu\nu}\mathcal{L}-2\nabla^{\alpha}\nabla^{\beta}P_{\mu\alpha\beta\nu}=0\, ,
\label{fieldequations}
\end{equation}
where we defined
$P^{\mu\nu\sigma\rho}\equiv \left[\partial \mathcal{L}/\partial R_{\mu\nu\sigma\rho}\right]|_{g_{\alpha\beta}}$.
Our aim is to linearize the equations \req{fieldequations} 
around a maximally symmetric spacetime (m.s.s.) $(\bar{\mathcal{M}},\bar{g}_{\mu\nu})$ with Riemann tensor
$
\bar R_{\mu\nu\alpha\beta}=2\Lambda \bar g_{\mu[\alpha}\bar g_{\beta]\nu}.
$
Hence, we consider a metric of the form $g_{\mu\nu}=\bar{g}_{\mu\nu}+h_{\mu\nu}$, 
where $h_{\mu\nu}\ll 1$ for all $\mu,\nu=0,\dots,D-1$. The goal is now to expand the field equations (\ref{fieldequations}) to linear order in the perturbation $h_{\mu\nu}$ and its derivatives. 
%By doing so, we obtain partial derivatives of the Lagrangian $\mathcal{L}$ both with respect to the metric and with respect to the Riemann tensor evaluated on the background. Without loss of generality, it is always possible to consider the Lagrangian as a function of the Riemann tensor with mixed indices $R^{\mu\nu}_{\ \ \alpha\beta}$ alone, which implies $\partial_{g_{\mu\nu}}\mathcal{L}=0$ \cite{Tekin2,Padmanabhan:2013xyr}. This means that all expressions will be characterized by the partial derivatives of the Lagrangian with respect to the Riemann tensor \footnote{Note also that, although formally we consider the Lagrangian to be a function of $R^{\mu\nu}_{\ \ \alpha\beta}$, there is no ambiguity in writing the derivatives with different index placements in the Riemann tensor.}.
In the process, it is convenient to define $\bar{\mathcal{L}}\equiv \mathcal{L}\big|_{B}$ and the objects
\begin{align}\label{P-def}
\bar P^{\mu\alpha\beta\nu}\equiv \frac{\partial\mathcal{L}}{\partial R_{\mu\alpha\beta\nu}}\Big|_{B}\, ,\quad \bar C_{\sigma\rho\lambda\eta}^{\mu\alpha\beta\nu}\equiv \frac{\partial P^{\mu\alpha\beta\nu}}{\partial R^{\sigma\rho\lambda\eta}}\Big|_{B}\, ,
\end{align}
where $\big|_B$ denotes evaluation on the background. 
%In terms of these, the variations of $\mathcal{L}$ and $P^{\mu\alpha\beta\nu}$ read
 %\begin{align}
%\delta\mathcal{L}&=\delta g^{\mu\nu} \bar P_{\mu}^{\ \sigma\rho\lambda}\bar R_{\nu\sigma\rho\lambda}+\bar P_{\mu}^{\ \sigma\rho\lambda}\delta R^{\mu}_{\ \ \sigma\rho\lambda}\, ,\\
%\delta P^{\mu\alpha\beta\nu}&=2\delta g^{\lambda[\mu}\bar P_{\lambda}^{\ \alpha]\beta\nu}+\delta g^{\rho\eta}\bar %C^{\mu\alpha\beta\nu}_{\lambda\eta\sigma\tau}\bar R^{\lambda \ \ \sigma\tau}_{\ \ \rho}\\ \notag &+\bar C^{\mu\alpha\beta\nu}_{\lambda\rho\sigma\tau}\bar g^{\rho \gamma}\bar g^{\sigma \kappa}\bar g^{\tau \upsilon}\delta R^{\lambda}_{\ \ \gamma \kappa \upsilon}\, .
%\end{align}
Now, the explicit form of tensors $\bar P^{\mu\alpha\beta\nu}$ and $\bar C_{\sigma\rho\lambda\eta}^{\mu\alpha\beta\nu}$ depends on the particular Lagrangian $\mathcal{L}$ under consideration. However, since $\bar P^{\mu\alpha\beta\nu}$ and $\bar C_{\sigma\rho\lambda\eta}^{\mu\alpha\beta\nu}$ are defined on a m.s.s., they can only contain terms involving products and contractions of the metric $\bar g_{\mu\nu}$, its inverse $\bar g^{\mu\mu}$ and $\delta_{\ \mu}^{\nu}$. Besides, as it is evident from their definitions \req{P-def}, they inherit the symmetries of the Riemann tensors involved in such definitions. 
These constrains are so strong that $\bar P^{\mu\alpha\beta\nu}$ and $\bar C_{\sigma\rho\lambda\eta}^{\mu\alpha\beta\nu}$ must be necessarily given by
 \begin{equation}
\begin{aligned}
\bar P^{\mu\alpha\beta\nu}&=2e \, \bar g^{\mu[\beta}\bar g^{\nu]\alpha}\, , \\
\bar C^{\sigma\rho\lambda\eta}_{\mu\alpha\beta\nu}&=a\left[\delta^{[\sigma}_{\mu}\delta^{\rho]}_{\alpha}\delta^{[\lambda}_{\beta}\delta^{\eta]}_{\nu}+\delta^{[\lambda}_{\mu}\delta^{\eta]}_{\alpha}\delta^{[\sigma}_{\beta}\delta^{\rho]}_{\nu}\right]\\  &+b\left[\bar g_{\mu\beta}\bar g_{\alpha\nu}-\bar g_{\mu\nu}\bar g_{\alpha\beta}\right]\left[\bar g^{\sigma\lambda}\bar g^{\rho\eta}-\bar g^{\sigma\eta}\bar g^{\rho\lambda}\right]\\
&+4\,c\,\delta^{[\sigma}_{\ (\tau}\bar g^{\rho][\lambda}\delta^{\eta]}_{\ \epsilon)}\delta^{\tau}_{\ [\mu}\bar g_{\alpha][\beta}\delta^{\epsilon}_{\ \nu]}\, ,
\end{aligned}
\label{abce-def}
\end{equation}
where the only theory-dependent quantities are the constants $a$, $b$, $c$ and $e$. Interestingly, these constants fully characterize the linearized equations of any theory of the form \req{Smass}.

Imposing the background metric $\bar g_{\mu\nu}$ to be a solution of the field equations \req{fieldequations} gives rise to a relation between the constant $e$ introduced in \req{abce-def}, the background scale $\Lambda$, and all the other possible couplings present in the general higher-order Lagrangian. It reads
$\bar{\mathcal{L}}(\Lambda)=4e(D-1)\Lambda$.
Using the chain rule and equations \req{abce-def} one can obtain another relation involving $e$ and $\Lambda$, namely
$d\bar{\mathcal{L}}(\Lambda)/d \Lambda=2eD(D-1)$.
These two equations together give rise to the beautiful expression
\begin{equation}
\label{Lambda-eq}
\Lambda\frac{d\bar{\mathcal{L}}(\Lambda)}{d \Lambda}=\frac{D}{2}\bar{\mathcal{L}}(\Lambda)\, .
\end{equation}
This is the relation between the scales and couplings appearing in the higher-derivative Lagrangian \req{Smass} and the background curvature $\Lambda$ which must be satisfied in order for $\bar g_{\mu\nu}$ to be a solution of \req{fieldequations}.

The information gathered so far is all what we need in order to linearize \req{fieldequations}. After a remarkably long computation, we obtain the following result for the linearized equations of \req{Smass}
%\begin{equation}
%\begin{aligned}
%\mathcal{E}_{\mu\nu}^{L}&=\left(2e+2\Lambda(2a+c(D-2))\right)G_{\mu\nu}^{ L}\\ \notag&+(2a+c)\big[2\left[\bar \Box-%D\Lambda  \right] G_{\mu\nu}^L-(D-2)\Lambda\bar g_{\mu\nu}R^L\big]\\
%&+2(2b+c+a)\left[\bar g_{\mu\nu}\bar\Box-\bar\nabla_{\mu}\bar\nabla_{\nu}\right]R^{ L}\\ \notag &+\Lambda(2a+D c %+4b (D-1))R^{ L}\bar g_{\mu\nu}
%\, .
%\end{aligned}
%\label{lineareqs}
%\end{equation}
\begin{equation}
\begin{aligned}
\frac{1}{2}\mathcal{E}_{\mu\nu}^{L}&=\left[e-2\Lambda(a(D-1)+c)+(2a+c)\bar \Box\right]G_{\mu\nu}^{ L}\\
&+\left[a+2b+c\right]\left[\bar g_{\mu\nu}\bar\Box-\bar\nabla_{\mu}\bar\nabla_{\nu}\right]R^{ L}\\ &-\Lambda\left[a(D-3)-2b(D-1)-c \right]R^{ L}\bar g_{\mu\nu}
\, ,
\end{aligned}
\label{lineareqs}
\end{equation}
where the linearized Einstein tensor, Ricci tensor and Ricci scalar read, respectively
\begin{align}
G_{\mu\nu}^L&=R_{\mu\nu}^L-\frac{1}{2}\bar g_{\mu\nu}R^L-(D-1)\Lambda h_{\mu\nu}\, ,\\
R_{\mu\nu}^L&=\bar\nabla^{\sigma}\bar\nabla_{(\mu}h_{\nu)\sigma}-\frac{1}{2}\bar\Box h_{\mu\nu}-\frac{1}{2}\bar\nabla_{\mu}\bar\nabla_{\nu}h\, , \\ 
R^{L}&=\bar\nabla^{\mu}\bar\nabla^{\nu}h_{\mu\nu}-\bar \Box h-(D-1)\Lambda h\, .
\end{align}
The problem is now that obtaining the values of $a$, $b$, $c$ and $e$ for a given theory using \req{P-def} and (\ref{abce-def}) is a rather tedious task in general, which involves computing first and second derivatives of the Lagrangian with respect to the Riemann tensor.
However, these parameters can actually be computed in a much simpler way for any theory.
Let us leave for the moment the framework of linearized gravity and consider an auxiliary symmetric tensor $k_{\mu\nu}$ whose indices --- as usual --- are raised with the inverse metric $g^{\mu\nu}$, and which by definition satisfies the following properties
\begin{equation}
\label{k-def}
k^{\mu}_{\ \mu}=\chi\, , \quad k^{\mu}_{\ \alpha}k^{\alpha}_{\ \nu}=k^{\mu}_{\ \nu}\, .
\end{equation} 
Here, $\chi$ is an arbitrary integer number smaller than $D$ that can be fixed at will, but which we will just leave undetermined throughout the calculation. Now let us define the following ``Riemann tensor'' %\footnote{The associated ``Ricci tensor'' and ``Ricci scalar'' are: $
%\tilde R_{\mu\nu}=\Lambda(D-1)g_{\mu\nu}+\alpha(\xi-1)k_{\mu\nu}$ and $\tilde R=\Lambda D(D-1)+\alpha \xi(\xi-1)$ respectively.
%}
\begin{equation}
\label{Riemalpha}
\tilde R_{\mu\nu\sigma\rho}(\Lambda, \alpha)=2\Lambda  g_{\mu[\sigma} g_{\rho]\nu}+2\alpha k_{\mu[\sigma}k_{\rho]\nu}\, , 
\end{equation}
where $\Lambda$ and $\alpha$ are two parameters. Observe that while $\tilde R_{\mu\nu\sigma\rho}(\Lambda, \alpha)$ has the right symmetries of a Riemann tensor, it is not --- or rather, it does not need to be --- the Riemann tensor of any actual metric in general. Note nevertheless that when $\alpha=0$, $\tilde R_{\mu\nu\sigma\rho}(\Lambda, \alpha)$ reduces to the Riemann tensor of a m.s.s. of curvature $\Lambda$. The next step is to evaluate our higher-derivative Lagrangian \req{Smass} on $\tilde R_{\mu\nu\sigma\rho}(\Lambda, \alpha)$, \ie we substitute all Riemann tensors appearing in $\mathcal{L}$ by the object defined in \req{Riemalpha}. This gives rise to a function of $\Lambda$ and $\alpha$,
\begin{equation}
\mathcal{L}(\Lambda,\alpha)= \mathcal{L}\left(R_{\mu\nu\rho\sigma}=\tilde R_{\mu\nu\rho\sigma}(\Lambda,\alpha),g_{\alpha\beta}\right)\, .
\label{Ldefinition}
\end{equation}
Note that in this evaluation, indices are still raised and lowered with $g_{\mu\nu}$, and not with some combination of $g_{\mu\nu}$ and $k_{\mu\nu}$. Now, it turns out that $a,b,c$ and $e$ can all be read off from partial derivatives of $\mathcal{L}(\Lambda,\alpha)$ with respect to $\alpha$ and evaluated at $\alpha=0$. 
%\footnote{Observe that we only need $\mathcal{L}(\Lambda,\alpha)$ up to $\alpha^2$ order, \ie from
%$
%\mathcal{L}(\Lambda,\alpha)=\mathcal{L}(\Lambda)+\left[2\chi(\chi-1)e\right]\, \alpha+ \left[ 2\chi(\chi-1)(a+b\,\chi(\chi-1)+c(\chi-1))\right]\, \alpha^2+\mathcal{O}(\alpha^3)
%$ we can read off the values of all the relevant constants.}. 
Such partial derivatives are of course trivial to compute. Indeed, using the chain rule along with equations (\ref{P-def}) and (\ref{abce-def}), we find that the following equations are satisfied for any theory,
\begin{align}\label{tete}
\frac{\partial \mathcal{L}}{\partial\alpha}\Big|_{\alpha=0}&=2e\,\chi(\chi-1)\,, \\ \label{teta}
\frac{\partial^2 \mathcal{L}}{\partial \alpha^2}\Big|_{\alpha=0}&=4\chi(\chi-1)\left(a+b\, \chi(\chi-1)+c(\chi-1)\right).
\end{align}
The crucial observation is that $a$, $b$, $c$ and $e$ do not depend on $\chi$. Hence, since these constants appear multiplied by factors involving different combinations of $\chi$, the above expressions allow us to identify them immediately for a given theory. Once we evaluate $\mathcal{L}(\Lambda,\alpha)$, we just need to compute the two partial derivatives above and compare the resulting expressions with the RHS of \req{tete} and \req{teta} in order to obtain $a$, $b$, $c$ and $e$. 

We have verified that our linearization procedure correctly accounts for all the results available in the literature corresponding to: quadratic gravities \cite{smolic,Lu,Tekin1,Tekin2,Deser:2011xc}, general $f($Lovelock$)$ theories \cite{Love} and QT gravity \cite{Quasi2,Quasi}.

Our method is remarkably simpler than computing $\bar{P}^{\mu\nu\rho\sigma}$ and $\bar C^{\mu\nu\alpha\beta}_{\lambda\eta\sigma\tau}$ explicitly using their definitions (\ref{P-def}). The hardest part of the calculation is the evaluation of the function $\mathcal{L}(\Lambda,\alpha)$, which exclusively involves trivial contractions of $g_{\mu\nu}$ and $k_{\mu\nu}$ and which, in the sense we have just explained, contains all the information necessary for the linearization of any theory of the form \req{Smass} on a m.s.s.. 

\textbf{Physical modes:} In order to identify the physical degrees of freedom propagated by the metric perturbation in a general theory of the form \req{Smass}, we further simplify the linearized equations \req{lineareqs} by imposing the transverse gauge condition, $\bar \nabla^{\mu} h_{\mu\nu}=\bar \nabla_{\nu} h$. A careful analysis suggests the following decomposition of the metric perturbation \footnote{Our analysis in this section is partially based on the one performed in \cite{smolic}.},
\begin{equation}
h_{\mu\nu}=t_{\mu\nu}^{(m)}+t_{\mu\nu}^{(M)}+\frac{[\bar\nabla_{\mu}\bar\nabla_{\nu}-\frac{1}{D}\bar g_{\mu\nu}\bar \Box]h}{(m_s^2+D\Lambda)}+\frac{1}{D}\bar g_{\mu\nu} h\, ,
\end{equation}
where $m_s^2$ will be defined in a moment. It is now possible to show that the traceless tensors $t_{\mu\nu}^{(m)}$ and $t_{\mu\nu}^{(M)}$, and the trace $h$ satisfy, respectively,
\begin{align}\notag
-\frac{1}{2\kappa_{\rm eff}}\left[\bar \Box -2\Lambda\right]t_{\mu\nu}^{(m)}&=0\, ,\\ \notag
\frac{1}{2\kappa_{\rm eff}}\left[\bar \Box -2\Lambda-m_g^2\right]t_{\mu\nu}^{(M)}&=0 , \\  
-\left[\frac{(D-1)(D-2)\Lambda(m_g^2-(D-2)\Lambda)}{4\kappa_{\rm eff}m_g^2(m_s^2+D\Lambda)}\right]\left[\bar \Box -m_s^2\right] h&=0\, , \label{cosisi}
\end{align}
where $\kappa_{\rm eff}$, $m_g^2$ and $m_s^2$ are defined in terms of $a$, $b$, $c$ and $e$ as
\begin{eqnarray}\label{kafka}
\kappa_{\rm eff}^{-1}&=&4e-8\Lambda(D-3)a\, ,\\ \label{kafka2}
m_g^2&=&(-e+2\Lambda(D-3)a)/(2a+c)\, ,\\ \label{kafka3}
m_s^2&=&\frac{e(D-2)-4\Lambda(a+(bD+c)(D-1))}{2a+Dc+4b(D-1)}\, .
\end{eqnarray}
Notice also that we have made explicit the overall coefficients which would appear in the LHS of equations \req{cosisi} had we coupled the theory to matter. In that case, an effective stress-tensor would appear in the RHS of equations \req{cosisi} \cite{PabloPablo}. From \req{cosisi}, it is clear that 
$t_{\mu\nu}^{(m)}$ is the usual massless graviton. $h$ is in turn a scalar field of mass $m_s$. Finally, $t_{\mu\nu}^{(M)}$ is a spin-2 field of mass $m_g$ whose coupling to matter would have the wrong sign, which is a manifestation of its ghost-like behavior. Also, $\kappa_{\rm eff}=8\pi G_{\rm eff}$ is the effective Einstein constant of the theory. Using \req{kafka}-\req{kafka3}, we can obtain the relevant physical parameters in terms of the constants characterizing a given theory as explained in the previous section. 

There exist different special classes of theories depending on the values of $m_s$ and $m_g$ --- see \eg \cite{Tekin1,Tekin2} for previous classifications. For example, if we demand $|m_g^2|=+\infty$, the massive graviton is absent, and the theory has a chance to be unitary. The family of $f($Lovelock$)$ gravities \cite{Love} is the prototypical example of this class --- including $f(R)$ \cite{Bueno2}, $R^2$ \cite{Tekin1,Alvarez-Gaume:2015rwa}, etc. If we set $|m_s^2|=+\infty$ instead, the scalar disappears and we are left with the two spin-2 fields. This is the case, \eg of quadratic conformal gravity \cite{conformal,Lu}. Within this class, we can further impose $m_g^2=0$, which leads to theories with two zero-energy gravitons. These are usually referred to as critical gravities \cite{Lu}. The case of theories with $|m_g^2|=|m_s^2|=+\infty$ will be treated in a moment.

\textbf{Linearization of cubic gravities:}
Let us now apply our linearization procedure to a general $D$-dimensional cubic gravity. The Lagrangian of such a theory can be written as
\begin{equation}
\begin{aligned}
\mathcal{L}=\frac{1}{2\kappa}\left[-2\Lambda_0+R\right]+\sum_{i=1}^{3}\alpha_i\mathcal{L}_i^{(2)}
+\kappa\sum_{i=1}^{8}\beta_i\mathcal{L}_i^{(3)}\, .
\end{aligned}
\label{quarti}
\end{equation}
In this expression, $\kappa\equiv 8\pi G$ and $\Lambda_0$ are the Einstein and cosmological constants respectively, $\alpha_i$ and $\beta_i$ dimensionless parameters, and $\mathcal{L}^{(2)}$ and $\mathcal{L}^{(3)}$ are complete sets of quadratic and cubic invariants, which we choose to be $\mathcal{L}^{(2)}=\{R^2,\, R_{\mu\nu}R^{\mu\nu},\, R_{\mu\nu\rho\sigma}R^{\mu\nu\rho\sigma}\}$ and $\mathcal{L}^{(3)}=\{\tensor{R}{_{\mu}^{\rho}_{\nu}^{\sigma}}\tensor{R}{_{\rho}^{\gamma}_{\delta}^{\beta}}\tensor{R}{_{\gamma}^{\mu}_{\delta}^{\nu}}, \, \tensor{R}{_{\mu\nu}^{\rho\sigma}}\tensor{R}{_{\rho\sigma}^{\gamma\beta}}\tensor{R}{_{\gamma\beta}^{\mu\nu}},\, \tensor{R}{_{\mu\nu\rho\sigma}}\tensor{R}{^{\mu\nu\rho}_{\gamma}}R^{\sigma\gamma},$ $ \tensor{R}{_{\mu\nu\rho\sigma}}\tensor{R}{^{\mu\nu\rho\sigma}}R,\, \tensor{R}{_{\mu\nu\rho\sigma}}\tensor{R}{^{\mu\rho}}\tensor{R}{^{\nu\sigma}}, \, R^{\nu}_{\mu} R_{\nu}^{\rho} R_{\rho}^{\mu}\, ,\,  R_{\mu\nu}R^{\mu\nu}R\, , \,R^3\}$ respectively. 

Following the linearization procedure explained in the first section, we obtain: $a=\alpha_3-\frac{\kappa \Lambda}{2}[3\beta_1-12\beta_2-2(D-1)\beta_3-2D(D-1)\beta_4 ]$, $b=\frac{\alpha_1}{2}+\frac{\kappa\Lambda}{2}[4\beta_4+\beta_5+2(D-1)\beta_7+3D(D-1)\beta_8 ]$, $c=\frac{\alpha_2}{2}+\frac{\kappa\Lambda}{2}[3\beta_1+4\beta_3+(2D-3)\beta_5+3(D-1)\beta_6+D(D-1)\beta_7 ]$ and $e=\frac{1}{4\kappa}+\Lambda[D(D-1)\alpha_1+(D-1)\alpha_2+2]+\frac{\kappa\Lambda^2}{2}\cdot [3(D-2)\beta_1+12\beta_2+6(D-1)\beta_3+6D(D-1)\beta_4+3(D-1)^2(\beta_5+\beta_6+D\beta_7+D^2\beta_8)]$.
%Using the method developed in --- which we review in the supplemental material --- we find that the linearized equations for this class of theories are given by
%\begin{equation}
%\begin{aligned}
%\mathcal{L}_1^{(2)}=R^2\,, \mathcal{L}_2^{(2)}=R_{ab}R^{ab}\, , \, \mathcal{L}_3^{(2)}=R_{abcd}R^{abcd}\, ,
%\end{aligned}
%\label{quarti}
%\end{equation}
%\begin{equation}
%\begin{aligned}
%\mathcal{E}_{\mu\nu}^{L}&=\left[\eta_1+\eta_2\bar \Box \right]G_{\mu\nu}^{ L}+\eta_3 \left[\bar g_{\mu\nu}(\bar\Box+\eta_4)-\bar\nabla_{\mu}\bar\nabla_{\nu}\right]R^{ L}
%\, ,
%\end{aligned}
%\label{lineareqs}
%\end{equation}
%and where $\eta_4=\Lambda (D-1)-\eta_2/\eta_3 (D-2)$ and
%\begin{equation}
%\begin{aligned}
%a&=\alpha_3-\frac{\kappa \Lambda}{2}\left[3\beta_1-12\beta_2-2(D-1)\beta_3-2D(D-1)\beta_4 \right]
%\, ,\\ 
%b&=\frac{\alpha_1}{2}+\frac{\kappa\Lambda}{2}\left[4\beta_4+\beta_5+2(D-1)\beta_7+3D(D-1)\beta_8 \right]
%\, ,\\ 
%c&=\frac{\alpha_2}{2}+\frac{\kappa\Lambda}{2}\left[3\beta_1+4\beta_3+(2D-3)\beta_5+3(D-1)\beta_6\right.\\  &\left.+D(D-1)\beta_7 \right]
%\, ,\\
%e&=\frac{1}{4\kappa}+\Lambda\left[D(D-1)\alpha_1+(D-1)\alpha_2+2\right]+\frac{\kappa\Lambda^2}{2}\cdot \\ &\left[3(D-2)\beta_1+12\beta_2+6(D-1)\beta_3+6D(D-1)\beta_4\right. \\& \left.+3(D-1)^2(\beta_5+\beta_6+D\beta_7+D^2\beta_8)\right]\, .
%\end{aligned}
%\label{lineareq456s}
%\end{equation}
Plugging these values in \req{lineareqs}, we obtain the linearized equations of the general cubic theory \req{quarti}.

%\comment{Imponiendo diferentes condiciones se pueden obtener teorias con diferentes propiedades. Por ejemplo}

%\begin{equation}
%\begin{aligned}
%\eta_1&=\frac{1}{2\kappa}+\Lambda\left[2D(D-1)\alpha_1+2(D-2)\alpha_2-4(D-2)\alpha_3\right]\\ \notag &+\kappa %\Lambda^2\left[9(D-2)\beta_1-12(2D-3)\beta_2\right.\\ \notag &-(4D(D-7D)+18)\beta_3-(2D(2D^2-7D+5))\beta_4\\ \notag &+(D(3D-10)+9)\beta_5+(3D(D-4)+9)\beta_6\\ \notag &\left.+(D(3D-8)+5)D\beta_7+3D^2(D-1)^2\beta_8\right]
%\, ,\\ 
%\eta_2&=\alpha_2+4\alpha_3+\kappa\Lambda\left[-3\beta_1+24\beta_2\right. \\ \notag & +4D\beta_3+4D(D-1)\beta_4 +(2D-3)\beta_5 \\ \notag &\left. +3(D-1)\beta_6+D(D-1)\beta_7\right]
%\, ,\\ 
%\eta_3&=2\alpha_1+\alpha_2+2\alpha_3+\kappa\Lambda\left[12\beta_2+2(D-1)\beta_3\right. \\ \notag &+(2D(D-1)+8)\beta_4  +(2D-1)\beta_5+3(D-1)\beta_6\\ \notag &\left.+(D(D+3)-4)\beta_7+6D(D-1)\beta_8\right]
%\, ,
%\end{aligned}
%\label{lineareqs}
%\end{equation}

\textbf{Einsteinian cubic gravity:} The theories which only propagate a massless graviton on the vacuum are those for which the additional modes are infinitely heavy, \ie those for which $|m_g^2|=|m_s^2|=+\infty$. These conditions translate into $2a+c=4b+c=0$. At quadratic order, they impose $\alpha_1=\alpha_3=-\alpha_2/4$. Hence, the only quadratic theory whose spectrum coincides with that of Einstein gravity is nothing but Gauss-Bonnet, $\mathcal{X}_4=R^2-4R_{\mu\nu}R^{\mu\nu}+R_{\mu\nu\rho\sigma}R^{\mu\nu\rho\sigma}$. For the cubic case, the two constraints would leave us with six independent parameters. Hence, there exists a six-parameter family of cubic theories whose spectrum is identical to that of Einstein gravity. Note however that most of those theories will be such that the values of the couplings $\beta_i$ will change with the spacetime dimension. Hence, they are actually \emph{different} theories in different dimensions. This is the case, for example, of QT gravity \cite{Quasi2,Quasi} and of certain $f($Lovelock$)$ theories \cite{Love,Tekin3}.

%Observe however that in most cases, these theories have dimension-dependent couplings, \ie the relative coefficients of the different curvature invariants change with the spacetime dimension. 

 Another aspect to be considered is the following. Consider a theory with dimension-dependent couplings $\mu(D)$ which only propagates the usual graviton in (A)dS$_D$, for some $D$. If, in the same number of dimensions, we linearize its equations on (A)dS$_{D^{\prime}}\times \mathbb{R}^{D-D^{\prime}}$ instead and consider configurations constant on the transverse factor, the equations will actually be identical to the ones that we would find for the theory --- with the same $\mu(D)$ couplings --- in $D^{\prime}$ dimensions. But these will in general involve the spin-2 ghost field. Of course, this problem does not occur if the theory is the same in all dimensions.

Interestingly, if we demand the parameters $\beta_i$ to be independent of $D$, the number of constraints is six instead of two. This is because both $2a+c$ and $4b+c$ are polynomials of order $D^2$, so we need to impose that the coefficients of the terms proportional to $D^0$, $D$ and $D^2$ vanish independently. This leaves us with a two-parameter family of theories. In particular, we find
\begin{align}
&\beta_1=\gamma,\, \beta_2=\frac{\gamma}{12}+\frac{14\zeta}{3},\, \beta_3=-24\zeta, \, \beta_4=3\zeta,\\ \notag
&\beta_5=-\gamma+16\zeta,\, \beta_6=\frac{2\gamma}{3}+\frac{64\zeta}{3}, \, \beta_7=-12\zeta\, ,\, \beta_8=\zeta\, ,
\end{align}
where $\gamma$ and $\zeta$ are the free parameters. Now we have the freedom to choose a basis of two cubic invariants satisfying the above constraints. The first element is somewhat canonical, and corresponds to the dimensionally-extended Euler density $\mathcal{X}_6$, which one finds for $\gamma=-8\zeta$.
%\begin{equation}
%\begin{aligned}
%\mathcal{X}_6=&-8R_{a\ b}^{\ c \ d}R_{c\ d}^{\ e \ f}R_{e\ f}^{\ a \ b}+4R_{ab}^{cd}R_{cd}^{ef}R_{ef}^{ab}-24R_{abcd}R^{abc}_{e}R^{de}+3R_{abcd}R^{abcd}R\\
%&+24R_{abcd}R^{ac}R^{bd}+
%16R_{a}^{b}R_{b}^{c}R_{c}^{a}-12R_{ab}R^{ab}R+R^3,
%\end{aligned}
%\end{equation}
 Any other choice produces another invariant. A particularly simple one corresponds to setting $\gamma=12$, $\zeta=0$, for which we get the following cubic term:
 \begin{align}\label{pseudo}
\mathcal{P}&=12 \tensor{R}{_{\mu}^{\rho}_{\nu}^{\sigma}}\tensor{R}{_{\rho}^{\gamma}_{\sigma}^{\delta}}\tensor{R}{_{\gamma}^{\mu}_{\delta}^{\nu}}+R_{\mu\nu}^{\rho\sigma}R_{\rho\sigma}^{\gamma\delta}R_{\gamma\delta}^{\mu\nu}\\ \notag &-12R_{\mu\nu\rho\sigma}R^{\mu\rho}R^{\nu\sigma}+8R_{\mu}^{\nu}R_{\nu}^{\rho}R_{\rho}^{\mu}\, .
\end{align}
 Hence, we find that the most general theory of the form \req{quarti} possessing dimension-independent couplings which shares spectrum with Einstein gravity reads
 \begin{equation}
\begin{aligned}
\mathcal{L}=\frac{1}{2\kappa}\left[-2\Lambda_0+R\right]+\alpha \mathcal{X}_4 + \kappa \left[\beta \mathcal{X}_6 + \lambda \mathcal{P} \right]\, .
\end{aligned}
\label{quartiii}
\end{equation}
 where $\mathcal{X}_4$ and $\mathcal{X}_6$ are the quadratic and cubic Lovelock terms respectively \footnote{We have cross-checked the linearized equations of motion of $\mathcal{P}$ for $D=4,5,6$ using the Mathematica package xAct \cite{xact}.}. The linearized equations of \req{quartiii} are almost identical to those of Einstein gravity, namely 
 \begin{equation}
\begin{aligned}
\mathcal{E}_{\mu\nu}^{L}&=\frac{1}{2\kappa_{\rm eff}}G_{\mu\nu}^{ L}\, ,
\end{aligned}
\label{lineareq44s}
\end{equation}
 the only signature of the higher-derivative theories being an effective Einstein constant given by $\kappa_{\rm eff}^{-1}=\kappa^{-1}[1+4(D-3)(D-4)\alpha (\Lambda \kappa)+6(D-3)(D-6)((D-4)(D-5)\beta-4\lambda)\Lambda^2\kappa^2]$.
 
 Observe that if we restrict ourselves to four dimensions, $\mathcal{X}_6=0$ identically, and $\mathcal{X}_4$ is topological. However, $\mathcal{P}$ is neither trivial nor topological in $D=4$. In fact, the embedding equation on a maximally symmetric spacetime \req{Lambda-eq} for this theory reads
  \begin{equation}\label{ems}
 8\lambda (D-3)(D-6)(\kappa\Lambda)^3-\kappa\Lambda+\frac{2\kappa\Lambda_0}{(D-1)(D-2)}=0\,.
 \end{equation}
 As we can see, $\mathcal{P}$ contributes to this equation in all spacetime dimensions but $D=3$ and $D=6$. \footnote{Note that the fact that these expressions reduce to those of Einstein gravity in $D=6$ follows from the fact that the contribution to the equations of motion of all six-dimensional cubic gravities vanishes identically for any Einstein metric.}. 
 Similarly, the effective Einstein constant reads $ \kappa_{\rm eff}^{-1}=\kappa^{-1}[1-24\lambda (D-3)(D-6)(\kappa\Lambda)^2]$ which, once the vacuum has been determined using \req{ems}, can be rewritten as $ \kappa_{\rm eff}^{-1}=\kappa^{-1}[6\Lambda_0/(\Lambda(D-1)(D-2))-2]$. The unitarity of the theory requires $\kappa_{\rm eff}$ to be positive, which in particular implies that the energy of the massless graviton is positive: $E=\kappa_{ E}/\kappa_{\rm eff}\cdot E_{ E}>0$, where the label $E$ stands for Einstein gravity \cite{Lu}. Using this condition, we observe that the allowed vacua are constrained to satisfy $0< (D-1)(D-2)\Lambda/(6\Lambda_0)<1/2$. For $D<6$, this puts a bound on the coupling value, $\lambda >-(D-1)^2(D-2)^2/(216(D-3)(6-D)\kappa^2\Lambda_0^2)$. If this is satisfied, then there is always at least one stable vacuum. If, in addition, $\lambda>0$, it follows from \req{ems} that this is unique.

\textbf{Conclusions:} We have notably simplified the problem of linearizing a general higher-order gravity of the form \req{Smass} on a m.s.s.. Our method only involves the evaluation of the gravitational Lagrangian on a particular Riemann tensor  $\tilde{R}_{\mu\nu\rho\sigma}(\Lambda,\alpha)$ --- defined in terms of two parameters in \req{Riemalpha} --- and of two derivatives of the resulting function with respect to one of the parameters. Previous methods ranged from the brute-force linearization of the corresponding full non-linear equations, to more refined techniques --- see \eg \cite{Tekin1,Tekin2,Tekin4} --- which incorporated decompositions analogous to the one performed in \req{abce-def}, but which still relied on the tedious computation of first and second derivatives of the Lagrangian with respect to the Riemann tensor. 
%In fact, using our method, we have been able to linearize all curvature invariants up to quartic order in curvature in general dimensions \cite{PabloPablo} in a reasonable time.

We have shown that, up to cubic order in curvature, there are only two theories with dimension-independent couplings which propagate a transverse, massless and positive-energy graviton on a m.s.s. and which are non-trivial in four dimensions. The first is of course Einstein gravity, and the second --- which we have coined \emph{Einsteinian cubic gravity} --- is given in \req{pseudo}. For positive values of the coupling constant $\lambda$, this theory has a unique vacuum. 

There are of course many aspects of the theory that have not been covered here and which should be studied elsewhere. %These include, for example, understanding the unitarity of the theory on more general backgrounds or the construction and study of cosmological and black hole solutions. 
For example, preliminary results suggest that it admits spherically symmetric black hole solutions of the form $ds^2=-f(r)dt^2+dr^2/g(r)+r^2d\Omega_{D-2}^2$, where $f(r)=g(r)$, \ie characterized by a single function.

We would also like to explore holographic applications --- \eg along the lines of \cite{Myers:2010jv,Buchel:2009sk}. For example, one of the three charges  --- $t_2$, $t_4$, $\ctt$ \cite{Osborn:1993cr,Buchel:2009sk} --- characterizing the dual CFT stress-tensor 3-point function vanishes for all holographic Lovelock theories, namely $t^{\text{Lovelock}}_4=0$. This partially motivated the construction of QT gravity \cite{Quasi}, for which $t_4$ is non-vanishing \cite{Myers:2010jv}, allowing for a $D\geq 5$ holographic model in which the three independent charges of the dual theory can be written in terms of three independent bulk quantities. In $D=4$ there are only two charges, as $t_2=0$ for all CFT$_3$ theories. However, QT gravity and Gauss-Bonnet --- for which $t_2\neq 0$ in $D\geq 5$ --- are trivial in that case, and Einstein gravity can only account for $\ctt$. Hence, it would be interesting to explore whether our theory can produce the remaining charge, $t_4$, in the AdS$_4$ case.

 More generally, it would be very desirable to understand to what extent the analogy with Einstein gravity provided here extends to other aspects of the theory. 

%Let us close by mentioning that there exist more \emph{psuedo-topological} gravities at higher orders in curvature. We will classify the quartic ones in a forthcoming publication \cite{PabloPablo}.\vspace{0.3cm}

\begin{acknowledgments} 
\emph{Acknowledgments:} We are thankful to Rob Myers and Tom\'as Ort\'in for useful comments. The works of PB and PAC were respectively supported by a postdoctoral fellowship from the Fund for Scientific Research Flanders (FWO) and a ``la Caixa-Severo Ochoa'' International pre-doctoral grant.
\end{acknowledgments}
\vspace{-0.7cm}

\bibliography{Cubic}

\end{document}